\documentclass[12pt]{article}

\usepackage{graphicx}
\usepackage{amsmath}
\usepackage{amssymb}
\allowdisplaybreaks

\begin{document}

\baselineskip=17.5pt plus 0.2pt minus 0.1pt

\renewcommand{\theequation}{\arabic{equation}}
\renewcommand{\thefootnote}{\fnsymbol{footnote}}
\makeatletter
\def\CR{\nonumber \\}
\def\pt{\partial}
\def\be{\begin{equation}}
\def\ee{\end{equation}}
\def\bea{\begin{eqnarray}}
\def\eea{\end{eqnarray}}
\def\eq#1{(\ref{#1})}
\def\la{\langle}
\def\ra{\rangle}
\def\hyp{\hbox{-}}


\begin{titlepage}
\title{\hfill\parbox{4cm}{ \normalsize YITP-09-22}\\
\vspace{1cm} Gauge fixing in the tensor model \\ 
and emergence of local gauge symmetries}
\author{
Naoki {\sc Sasakura}\thanks{\tt sasakura@yukawa.kyoto-u.ac.jp}
\\[15pt]
{\it Yukawa Institute for Theoretical Physics, Kyoto University,}\\
{\it Kyoto 606-8502, Japan}}
\date{}
\maketitle
\thispagestyle{empty}
\begin{abstract}
\normalsize
The tensor model can be regarded as theory of dynamical fuzzy spaces, and gives a 
way to formulate gravity on fuzzy spaces.
It has recently been shown that the low-lying fluctuations around the Gaussian background
solutions in the tensor model 
agree correctly 
with the metric fluctuations on the flat spaces with general dimensions in the general relativity.
This suggests that the local gauge symmetry (the symmetry of local translations) is
also emergent around these solutions. 
To systematically study this possibility, I apply the BRS gauge fixing procedure to the tensor model.
The ghost kinetic term is numerically analyzed, and it has been found that there exist
some massless trajectories of ghost modes, which are clearly separated from the other higher ghost modes. 
Comparing with the corresponding BRS gauge fixing 
in the general relativity, these ghost modes forming the massless trajectories in
the tensor model 
are shown to be identical to the reparametrization ghosts in the general relativity.  
\end{abstract}
\end{titlepage}

\section{Introduction}
\label{sec:intro}
Various thought experiments considering quantum gravitational fluctuations
have shown that the classical concept of smooth spacetime in the general relativity is not appropriate 
in some extreme cases \cite{Garay:1994en,Sasakura:1999xp}, 
and should be replaced in some way by a novel concept of quantum 
spacetime.
Fuzzy space\footnote{In this paper, this term is used as its widest meanings. It includes noncommutative 
spaces as well as nonassociative ones \cite{Sasai:2006ua}.}
 is one of such candidates of quantum space \cite{Connes:1994yd,Madore:2000aq,Balachandran:2005ew}. 
A fuzzy space is defined by an algebra of functions on it, 
unlike a classical spacetime being described by a coordinate system. 
This kind of algebraic definition of spaces has some physically interesting advantages 
over the classical description. For example, in quantum gravity, 
the changes of topologies and dimensions of space are believed to be the vital
processes of quantum fluctuations. However, it is generally hard or tightly constrained 
to describe these processes without encountering 
singularities in the classical description \cite{Carlip:1998uc}. On the contrary,
in general, one will have much more freedom to describe such processes
in fuzzy spaces  
through interpolation  
between algebraic structures of fuzzy spaces approximating 
classical spaces with distinct topologies and/or dimensions.
This kind of thoughts suggest an interesting research direction;
considering theory of dynamical fuzzy spaces as a model of quantum gravity.

In the recent years, there have been numerous discussions about gravity on fuzzy spaces.
A class of approaches discuss analogues of the general relativity on fuzzy spaces.
In this class of approaches, however, the dynamical variable is a fuzzy analogue of the 
metric tensor, and a fuzzy space itself is assumed to be fixed.
Therefore these approaches do not take the full advantages of the notion of fuzzy space
as explained above. 
On the contrary, a more interesting kind of approaches were initiated by 
the matrix models \cite{Banks:1996vh,Ishibashi:1996xs}.
These approaches consider spaces as dynamical objects generated as classical solutions or vacua,
and fluctuations of matrices around such vacua are regarded as field fluctuations 
on background fuzzy spaces.  
Then an extremely interesting possibility is that gravity may appear as one of these emergent
fields. So far this is yet an open issue under active 
investigations \cite{Kawai:2007zz,Steinacker:2009mb}.

In view of this present status, it might be meaningful to study another kind of model of 
dynamical fuzzy spaces, which is similar to but distinct from the matrix models. 
The model studied in this paper is the tensor model,
which has a rank-three tensor as its dynamical variable, instead of matrices in the matrix models. 
The tensor model was originally proposed to describe
the simplicial quantum gravity in dimensions greater 
than two \cite{Ambjorn:1990ge,Sasakura:1990fs,Godfrey:1990dt,
Boulatov:1992vp,Ooguri:1992eb,DePietri:1999bx,
DePietri:2000ii}\footnote{A tightly related kind of models, called the group field theories, 
have been being discussed mainly in the context of the loop quantum gravity. 
It is known that a certain group field theory can be considered to be a field theory 
on a noncommutative spacetime \cite{Imai:2000kq} and can also be derived as effective field 
theory of three-dimensional quantum gravity \cite{Freidel:2005me,Sasai:2009az}. 
See also \cite{Oriti:2009nd} for more and the recent developments.}.
The tensor model has not yet been successful 
for the analysis of the simplicial quantum gravity itself, 
in part because of the absence of the analytical methods to solve the tensor model.
However, it was recently proposed by the present author that the tensor model may be reinterpreted
as theory of dynamical fuzzy spaces \cite{Sasakura:2005js,Sasakura:2005gv,Sasakura:2006pq}. 
This is based on the fact that a fuzzy space can be 
characterized by a rank-three tensor $C_{ab}{}^c$ which determines the algebraic relations among 
all the functions $f_a$ on a fuzzy space 
through the product $f_a \star f_b=C_{ab}{}^cf_c$. From this point of view, 
it may not be necessary to analytically solve the tensor model to make relations to physics.
In analogy with the matrix model mentioned above,
a classical solution in the tensor model may be regarded as a background fuzzy space,
and the fluctuations of the tensor about it as field fluctuations. 
Then the question is whether gravity appears in such fluctuations. In fact, 
in a class of tensor models which have the classical solutions with Gaussian forms,
it has been shown that 
the low-lying fluctuations about these solutions 
at low momenta match correctly 
with the metric fluctuations on flat spaces in the general relativity in general dimensions
\cite{Sasakura:2007sv,Sasakura:2007ud,Sasakura:2008pe}.     

The above agreement is very interesting, but it is merely classical and obviously not enough 
quantum mechanically. The main purpose of this paper is to show the agreement a step further to
include the gauge degrees of freedom, which are the local translations in the present case. 
The tensor model has the symmetry of the orthogonal group $O(N)$, where $N$ is the 
number of all the functions or more physically ``points" forming a fuzzy space. 
A background solution of the tensor model
breaks this $O(N)$ symmetry down to some remaining symmetries of the solution or the background space, 
and the broken symmetries are 
realized non-linearly around it\footnote{ 
These modes of broken symmetries appeared as vanishing spectra of fluctuations
in the previous works \cite{Sasakura:2007sv,Sasakura:2007ud}.}. 
Since the broken symmetries permute the ``points" of the background fuzzy space,
they are intrinsically local symmetries, 
and it is tempting to insist that these are emergent local gauge symmetries 
(the local translation symmetry)\footnote{The idea to consider local gauge symmetries to be  
non-linearly realized broken symmetries is rather old. For example, 
see \cite{Ferrari:1971at,Brandt:1974jw,Borisov:1974bn}.}.
In this paper, 
to make this statement more precise and systematic, 
I will apply the BRS gauge fixing procedure to the tensor model and numerically
analyze the ghost kinetic term. Then I will compare the results of the numerical analysis
with the corresponding BRS gauge fixing in the general relativity.

This paper is organized as follows. In the following section, I will 
apply the BRS gauge fixing procedure to the tensor model. 
In Sec.\ref{sec:BRSrelativity}, I will discuss the corresponding BRS gauge fixing
procedure in the general relativity.
In Sec.\ref{sec:numanaly}, I will numerically study the eigenvalues and eigenmodes of 
the ghost kinetic term in the tensor model at the Gaussian backgrounds with 
dimensions $D=1,2,3$, 
and compare with the ghost kinetic term in the general relativity
on the flat spaces in these dimensions.
The final section is devoted to a summary and discussions.

\section{BRS gauge fixing procedure in the tensor model}
\label{sec:BRStensor}

\subsection{Direct computation of the gauge volume}
\label{sec:direct}
Let me start with the direct computation of the gauge volume.

The dynamical variable of the tensor model in this paper 
is given by a real rank-three tensor 
$C_{abc}$, which is totally symmetric,
\be
C_{abc}=C_{bca}=C_{cab}=C_{bac}=C_{acb}=C_{cba}.
\ee
The index takes values $1,2,\ldots,N$.
There is also a {\it nondynamical} symmetric real tensor $g^{ab}$, which is 
basically taken to be $g^{ab}=\delta^{ab}$.
Therefore,
by following the standard pairwise index contractions, the tensor model 
is invariant under the orthogonal group transformation $O(N)$,
\be
C_{abc}\rightarrow (MC)_{abc}\equiv M_a{}^{a'} M_b{}^{b'} M_c{}^{c'}C_{a'b'c'},
\ee 
where $M_{a}{}^{a'}\in O(N)$. This is the gauge symmetry of the tensor model.

The $O(N)$ symmetric metric in the space of the dynamical variable $C$ 
is defined by\footnote{ 
There exists an ambiguity to add $dC_{ab}{}^{b}dC^{ac}{}_c$ to this metric.
The addition will change some details of 
the analysis of both the tensor model and the continuum theory, but
the mutual agreement should be obtained anyway.}
\be
\label{eq:cmeasure}
ds_C^2=dC_{abc}dC^{abc}.
\ee 
The inner product associated with the metric \eq{eq:cmeasure} between 
two rank-three totally symmetric tensors is defined by
\be
\langle A, B \rangle=A_{abc}B^{abc}.
\ee

The infinitesimal $SO(N)$ transformation of $C$ is given by
\be
(T^i C)_{abc}\equiv T^i{}_{a}{}^{a'} C_{a'bc}+T^i{}_b{}^{b'}C_{ab'c}+T^i{}_c{}^{c'}C_{abc'},
\ee
where $T^i{}_a{}^{a'}$ ($i=1,2,\ldots,N(N-1)/2$)
are the real antisymmetric matrices forming the Lie-algebra $so(N)$ in the vector 
representation.

The volume measure in the space of $C$ is defined from the metric \eq{eq:cmeasure}.
Dividing an infinitesimal region into the gauge directions and the others, 
the infinitesimal volume $dV_C$ can be expressed as 
\be
\label{eq:dvcdg}
dV_C=dg  \, dV_C^{\perp}\, \sqrt{\frac{\hbox{Det}(\langle TC,TC\rangle)}
{\hbox{Det}(\langle T,T \rangle)}},
\ee
where $dg$ is the Haar measure of $SO(N)$ and $dV_C^\perp$ denotes
the infinitesimal volume normal to the gauge directions.
Here Det($\cdots$) are the determinants of the matrices with components, 
\bea
\langle TC,TC\rangle_{ij}&=&\langle T^iC, T^jC\rangle, \\
\langle T,T\rangle_{ij}&=& h_0 \hbox{Tr}(T^i T^j),
\eea 
where Tr denotes the trace in the vector representation, and $h_0$ is a coefficient
related to the normalization of the Haar measure.
Since the integrand in \eq{eq:dvcdg} is invariant along the $SO(N)$ gauge directions,
the partial integration over the gauge directions is trivially performed as 
\be
\int_{SO(N)} dV_C=dV_C^{\perp}\, \frac{\hbox{Vol}\left(O(N)\right)}{n}
\sqrt{\frac{\hbox{Det}(\langle TC,TC\rangle)}
{\hbox{Det}(\langle T,T \rangle)}},
\ee 
where $n$ is the possible symmetry factor becoming larger than 1 if there exists a non-trivial 
$M\in SO(N)$ which satisfies $C=MC$.  
This factor $n$ can practically be ignored,
since the regions of such symmetric values of $C$
have generally vanishing volumes in the space of $C$. 
Thus, ignoring all the factors independent of $C$, one finally obtains
\be
\label{eq:dvc}
\int_{SO(N)} dV_C=dV_C^{\perp}\, \sqrt{\hbox{Det}(\langle TC,TC\rangle)}.
\ee

\subsection{BRS gauge fixing procedure in the tensor model}
I apply the general BRS gauge fixing scheme with the so-called $B$ field
presented in \cite{Kugo:1981hm} to the $SO(N)$ symmetry in the tensor model.
The BRST transformation of $C$ is given by
\be
(\delta_B C)_{abc}=c_i(T^i C)_{abc},
\ee
where $c_i$ are the ghost variables, which are assumed to be real.
The BRST transformation of the ghost variables is given by
\be
\delta_B c_k=\frac12 f^{ij}{}_k c_i c_j,
\ee
where $f^{ij}{}_k$ is the structure constant of $so(N)$, defined by $[T^i,T^j]=f^{ij}{}_k T^k$. 
There are also the anti-ghost and the B-variables, the BRST 
transformations of which are given by
\bea
\delta_B \bar{c}_i&=&i B_i. \cr
\delta_B B_i&=&0. 
\eea
These $\bar c_i$ and $B_i$ are also assumed to be real.
The nilpotency $\delta_B^2=0$ can easily be shown by explicit computations. 

The interest of the present paper is in the small fluctuations around certain backgrounds of $C$.
Let me denote a background by $C^0$ and the fluctuations by $A$,
\be
C_{abc}=C^0_{abc}+A_{abc}.
\ee
Then the dynamical variable is shifted to $A$, and its BRST transformation is given by
\be
(\delta_B A)_{abc}=(\delta_B C)_{abc}=c_i(T^i C)_{abc}=c_i(T^i C^0)_{abc}+c_i(T^iA)_{abc}.
\ee

The general scheme implies that the BRST exact action corresponding to the sum of the Faddeev-Popov 
and the gauge fixing terms can generally be given by
\be
\label{eq:fppgf}
S_{GF+FP}=\delta_B\left(\bar c_i F^i(A,c,\bar c,B)\right),
\ee
where $F^i$ are the (almost arbitrary) gauge-fixing functions with vanishing ghost number. 
A natural choice in the present case is
\be
\label{eq:fi}
F^i=\langle T^iC^0,A\rangle,
\ee
since the gauge fixing conditions ($F^i=0$) only allow $A$ to be 
normal to the gauge directions at the background $C^0$. 
Computing \eq{eq:fppgf} with \eq{eq:fi}, $S_{GF+FP}$ is explicitly given by 
\be
\label{eq:expfpgf}
S_{GF+FP}=i B_i \langle T^iC^0,A\rangle- \bar c_i \langle T^iC^0,T^j C \rangle c_j.
\ee 

The path integral measure, which is just a usual integration in the present case, can be defined 
by
\be
\label{eq:BRSvolmeasure}
\int [dA]\prod_i dB_i dc_i d\bar c_i,
\ee
where $[dA]$ is the volume measure of $A$ defined from the metric $ds^2_A=dA_{abc}dA^{abc}$,
which is identical to the $O(N)$ symmetric metric \eq{eq:cmeasure}. 
Here a possible overall factor is not taken care of.
From the $O(N)$ invariance of the volume measure $[dA]$, 
one can easily prove the BRST invariance of the integral,
\be
\int [dA]\prod_i dB_i dc_i d\bar c_i\ \delta_B(\ldots)=0,
\ee
which guarantees the independence of physics from the choice of the gauge-fixing functions.

\subsection{Comparison between the direct and the BRS expressions}
In the following, let me compare the BRS result \eq{eq:expfpgf}, \eq{eq:BRSvolmeasure}
with the direct computation \eq{eq:dvc}.
To do this, let me introduce a normalized orthogonal basis which divides the space about $C^0$
into the subspaces tangent $\{v^{0||}_i\}$ and normal $\{v_l^{0\perp}\}$ to the gauge directions,
\bea
\label{eq:c0basis}
\langle T^i C^0,v_l{}^{0\perp}\rangle&=&0,\cr
\langle v^{0\perp}_l,v^{0\perp}_m \rangle&=&\delta_{lm},\cr
\langle v^{0\perp}_l,v^{0||}_i \rangle&=&0, \cr
\langle v^{0||}_i,v^{0||}_j\rangle&=&\delta_{ij}.
\eea
In general, $A$ can be expanded in terms of these vectors as
\be
\label{eq:Aexpand}
A=\alpha^i v^{0||}_i+\beta^l v^{0\perp}_l.
\ee
From the definition of the basis \eq{eq:c0basis}, the volume measure is
$[dA]=\prod_{i} d\alpha^i \prod_l d\beta^l$.
Putting \eq{eq:Aexpand} into \eq{eq:expfpgf}, and integrating over 
$c_i,\bar c_i,B_i$ and finally over $\alpha^i$,
one obtains
\be
\label{eq:FPexp}
\int [dA]\prod_i dB_i dc_i d\bar c_i e^{-S_{GF+FP}-S(C)}=
\int \prod_l d\beta^l \ \frac{\hbox{Det}(\langle TC^0,TC \rangle)}{\left| 
\hbox{Det}(\langle TC^0,v^{0||}\rangle)\right|} e^{-S(C^0+\beta^l v^{0\perp}_l)}, 
\ee
where an overall numerical constant is ignored, 
$S(C)$ is the original unfixed action, and the matrices in the determinants are defined by
\bea
\label{eq:tcmatrix}
\langle TC^0,TC\rangle_{ij}&=&\langle T^iC^0,T^jC\rangle,\cr
\langle TC^0,v^{0||}\rangle_{ij}&=&\langle T^iC^0,v_j^{0||}\rangle.
\eea

The result \eq{eq:FPexp} does not look like \eq{eq:dvc}, but they are actually identical.
To see this, let me introduce a similar normalized orthogonal basis $\{v^{||}_i\}$, $\{v_l^{\perp}\}$
around $C=C^0+A$ as \eq{eq:c0basis},
\bea
\label{eq:cbasis}
\langle T^i C,v_l{}^{\perp}\rangle&=&0,\cr
\langle v^{\perp}_l,v^{\perp}_m \rangle&=&\delta_{lm},\cr
\langle v^{\perp}_l,v^{||}_i \rangle&=&0, \cr
\langle v^{||}_i,v^{||}_j\rangle&=&\delta_{ij}.
\eea
Then the square of the determinants in \eq{eq:FPexp} can be computed as
\bea
\label{eq:squareFP}
\left(\frac{\hbox{Det}(\langle TC^0,TC \rangle)}{\left| 
\hbox{Det}(\langle TC^0,v^{0||}\rangle)\right|}\right)^2&=&
\hbox{Det}\left(\langle TC,TC^0 \rangle \left(\langle TC^0,v^{0||}\rangle 
\langle v^{0||},TC^0\rangle\right)^{-1} \langle TC^0,TC\rangle \right) \cr
&=& \hbox{Det}\left( \langle TC,v^{0||}\rangle \langle v^{0||},TC\rangle \right) \cr
&=& \hbox{Det}\left( \langle TC,TC\rangle\right) 
\left[\hbox{Det}\left(\langle v^{||},v^{0||}\rangle\right)\right]^2 \cr
&=& \hbox{Det}\left( \langle TC,TC\rangle\right) 
\left[\hbox{Det}\left(\langle v^{\perp},v^{0\perp}\rangle\right)\right]^2,
\eea
where similar shorthand notations like \eq{eq:tcmatrix} are used to denote the matrices.
In the above derivation, I have used the completeness of the bases 
$\left\{v^{||}\right\},\left\{ v^{0||}\right\}$ in the subspaces tangent to the gauge directions, and 
\be
\left[\hbox{Det}\left(\langle v^{||},v^{0||}\rangle\right)\right]^2=\hbox{Det}\left( 
\begin{array}{cc}
\langle v^{||},v^{||} \rangle & \langle v^{||},v^{0\perp} \rangle \\
\langle v^{0\perp},v^{||}\rangle & \langle v^{0\perp},v^{0\perp}\rangle
\end{array}
\right)=\left[\hbox{Det}\left(\langle v^{\perp},v^{0\perp}\rangle\right)\right]^2,
\ee
which can be shown from the identity,
\be
\hbox{Det}(D) \hbox{Det}(A-B D^{-1} C)=
\hbox{Det}\left(
\begin{array}{cc}
A & B \\
C & D
\end{array}
\right)= \hbox{Det} (A) \hbox{Det}(D-C A^{-1} B),
\ee
and the properties of the orthogonal normalized bases. 
From the definition \eq{eq:Aexpand} and that $dV_C^\perp$ in \eq{eq:dvc} is the infinitesimal
volume normal to the gauge directions at $C$, one obtains
\be
\label{eq:jacobi}
dV^\perp_C=\prod_l d\beta_l \left| \hbox{Det}\left(\langle v^\perp,v^{0\perp}\rangle\right)\right|.
\ee
Thus \eq{eq:dvc} and \eq{eq:FPexp} are actually identical, because of \eq{eq:squareFP}, \eq{eq:jacobi}.
 
\section{BRS gauge fixing in the general relativity}
\label{sec:BRSrelativity}
In this subsection, I will discuss the BRS gauge fixing procedure in the general 
relativity \cite{Nakanishi:1977gt,Nishijima:1978wq,
Kugo:1978rj,Nakanishi:1990qm}
corresponding
to that of the tensor model in the previous section.

By rewriting the coordinate transformation of the metric tensor with the ghost fields,
the BRST transformation of the metric tensor is given by
\be
\delta_B g_{\mu\nu}=\nabla_\mu c_\nu+\nabla_\nu c_\mu,
\ee
where $\nabla_\mu$ is the covariant derivative, and $c_\mu$ is the ghost vector field.
Then the nilpotency of the BRST transformation requires that 
the ghost field be transformed by
\bea
\delta_B c_\mu&=&-c^\nu \nabla_\mu c_\nu, \cr
(\delta_B c^\mu&=&c^\nu\nabla_\nu c^\mu=c^\nu\partial_\nu c^\mu). 
\eea
The anti-ghost field and the $B$-field are
introduced with the BRST transformations,
\bea
\delta_B \bar c_\mu&=&i B_\mu, \cr
\delta_B B_\mu&=&0.
\eea
The nilpotency $\delta_B^2=0$ can be checked by explicit 
computations\footnote{For example, $\delta_B^2 c_\mu=0$ can be shown from 
$\delta_B \Gamma_{\mu\nu}{}^\rho=c^\sigma R_{\mu\sigma\nu}{}^\rho+\nabla_\mu\nabla_\nu c^\rho$ and
the Bianchi identities for the Riemann tensor.}.
All the fields above are assumed to be real.

In the numerical analysis of the following section, I will take $C^0$ to be the Gaussian 
backgrounds \cite{Sasakura:2007sv,Sasakura:2007ud,Sasakura:2007ud}, 
which correspond to the fuzzy flat spaces with arbitrary dimensions. 
Correspondingly, flat backgrounds are considered in the general relativity as  
\be
g_{\mu\nu}=\delta_{\mu\nu}+h_{\mu\nu},
\ee
where $h_{\mu\nu}$ is the new dynamical field with 
$\delta_B h_{\mu\nu}=\nabla_{\mu}c_\nu+\nabla_\nu c_\mu$.
 
In \cite{Sasakura:2007ud}, it was argued that the metric \eq{eq:cmeasure} corresponds 
to the DeWitt supermetric \cite{DeWitt:1962ud},
\be
\label{eq:dewitt}
ds^2_g=\int d^Dx \sqrt{g}  (g^{\mu\nu}g^{\rho\sigma}+4 g^{\mu\rho}g^{\nu\sigma}) \delta g_{\mu\nu} 
\delta g_{\rho\sigma}.
\ee
Thus the inner product associated to \eq{eq:dewitt} between two
rank-two symmetric tensor fields is defined by
\be
\langle k,l \rangle_g=\int d^Dx 
\sqrt{g}  (g^{\mu\nu}g^{\rho\sigma}+4 g^{\mu\rho}g^{\nu\sigma}) 
k_{\mu\nu} l_{\rho\sigma}.
\ee

In the following, I want to obtain the BRS gauge fixing in the general relativity 
which is analogous to \eq{eq:fppgf}, \eq{eq:fi}.
Since $C^0$ corresponds to the flat background, the analogy of $T^i C^0$ are 
the infinitesimal local translations of the flat background. Therefore 
$\bar c_i T^i C^0$ in the tensor model should correspond 
to the field $\partial_\mu \bar c_\nu + \partial_\nu \bar c_\mu$. 
The deviation $A$ from the background in the tensor model corresponds to $h_{\mu\nu}$.
Thus the action corresponding to \eq{eq:fppgf}, \eq{eq:fi} 
is obtained as
\bea
\label{eq:gfpfpg}
S^g_{GF+FP}&=&\delta_B \langle \partial_\mu \bar c_\nu+\partial_\nu \bar c_\nu, h_{\rho\sigma}\rangle_g \cr
&=& 2\ \delta_B \left(\int d^Dx \sqrt{g}(g^{\mu\nu}g^{\rho\sigma}+4 g^{\mu\rho}g^{\nu\sigma})
(\partial_\mu \bar c_\nu) h_{\rho\sigma}\right).
\eea
In the quadratic order of the fields around the flat background, 
\eq{eq:gfpfpg} becomes 
\bea
\label{eq:gf2}
S^{g(2)}_{GF+FP}= 2 \int d^Dx\left[  
i\,(\partial^\mu B_\mu\, h^\nu_\nu+4\, \partial_\mu B_\nu\, h^{\mu\nu})
-6\, \partial^\mu \bar c_\mu \, \partial^\nu c_\nu-4\, \partial_\mu \bar c_\nu\, \partial^\mu c^\nu
\right].
\eea

The partial derivative of \eq{eq:gf2} with respect to $B_\mu$ gives the gauge fixing condition as
\be
\partial_\mu h^\nu_\nu+4 \partial^\nu h_{\mu\nu}=0.
\ee 
One can check that 
this gauge fixing condition is actually satisfied by all the metric fluctuation modes 
corresponding to the low-lying fluctuation modes in the tensor model
which were reported previously in \cite{Sasakura:2007ud,Sasakura:2008pe}.

Putting the form $c_\mu,\bar c_\mu = n_\mu e^{ipx}$,
the kinetic term of the ghost fields in \eq{eq:gf2} can be shown to have the spectra,
\be
\label{eq:spectracon}
\left\{
\begin{array}{ll}
20\, p^2 & \hbox{for the longitudinal mode\ } n_\mu \propto p_\mu ,\\
8\, p^2 &  \hbox{for the normal modes\ } n_\mu p^\mu=0.
\end{array}
\right.
\ee
Thus the longitudinal mode has no degeneracy, while the normal modes have the degeneracy $D-1$ in
$D$ dimensions. 

\section{Numerical analysis of the ghost kinetic term in the tensor model}
\label{sec:numanaly}
In the papers \cite{Sasakura:2007sv,Sasakura:2007ud,Sasakura:2008pe}, 
a class of tensor models possessing the classical solutions with Gaussian forms have 
been constructed and analyzed. 
In this section, I will take $C^0$ to be such Gaussian backgrounds, and numerically study the 
ghost kinetic term in the tensor model.
In all the dimensional cases to be studied ($D=1,2,3$), 
some massless trajectories of ghost modes, which are 
clearly separated from the other higher ghost modes, will be found, and 
they will be identified with the reparametrization ghosts
in the general relativity.

\subsection{The coefficient matrix of the ghost kinetic term}
In the momentum basis, such a gaussian background $C^0$ has the 
form \cite{Sasakura:2007sv,Sasakura:2007ud,Sasakura:2008pe,Sasai:2006ua}
\be
\label{eq:c0p}
C^0_{p_1,p_2,p_3}=\exp \left( -\alpha (p_1^2+p_2^2+p_3^2)\right) \delta_{p_1+p_2+p_3,0},
\ee
where $\alpha$ is a positive parameter, and 
each momentum is assumed to take integer vales bounded by $L$:
\be
\label{eq:pregion}
p=(p^1,p^2,\ldots,p^D),\ \ \ \ \sum_{i=1}^D (p^i)^2 \leq L^2.
\ee 
Since the delta function in \eq{eq:c0p} implies the momentum conservation,
there remains the $D$-dimensional translational symmetry on this background.
 
From \eq{eq:expfpgf}, the ghost kinetic term is given by
\be
\label{eq:sgh}
S_{gh}=-\bar c_i \langle T^iC^0,T^jC^0\rangle c_j.
\ee
Because of the momentum conservation of the background \eq{eq:c0p},
it is most convenient to take a momentum basis for the generators $T^i$.
Namely, the indices of the generators are given by pairs of distinct momenta $i=[p\,q]\ (p\neq q)$,
and the generators are antisymmetric matrices defined by
\be
\label{eq:tpq}
(T^{[p\,q]})_{rs}=\delta_{p,r}\delta_{q,s}-\delta_{p,s}\delta_{q,r}.
\ee
Then, putting \eq{eq:tpq} into \eq{eq:sgh},  
the coefficient matrix of the ghost kinetic term is given by
\bea
M^{gh}_{[p_1q_1],[p_2q_2]}&\equiv& \langle T^{[p_1q_1]}C^0,T^{[p_2q_2]}C^0\rangle \cr
&=&
3 \delta_{p_1+p_2,0}\delta_{q_1+q_2,0} 
\sum_{r,s}C^0_{q_1, r, s}C^0_{q_2,-r,-s}+
6 \sum_{r} C^0_{p_1,q_2,r}C^0_{q_1,p_2,-r} \cr
&&\ 
-(p_1\leftrightarrow q_1)-(p_2\leftrightarrow q_2)+(p_1\leftrightarrow q_1,p_2\leftrightarrow q_2),
\eea  
where the summations of the momenta $r,s$ are over the range \eq{eq:pregion}, 
and the simplified notations for anti-symmetrization have been used.

Because of the momentum conservation of the background, the matrix 
$M^{gh}_{[p_1q_1],[p_2q_2]}$ is divided into 
the block matrices with each value of the ghost momentum $p_1+q_1=-(p_2+q_2)$. 
Therefore, the analysis of $M^{gh}$ can be performed independently at each momentum sector.

In the following subsections, I will study the spectra and the properties of the eigenmodes of $M^{gh}$,
for dimensions $D=1,2,3$, and compare with the continuum theory.

The numerical facility was a Windows XP 64 workstation containing two Opteron 275 processors
and 8 GB memories. The C++ codes\footnote{The codes are downloadable from
http://www2.yukawa.kyoto-u.ac.jp/\~{}sasakura/codes/ghostcpp.zip.} 
were compiled by Intel C++ compiler 10.1
with ACML 4.2 for Lapack/Blas routines. The output were analyzed in Mathematica 6.0.

\subsection{The D=1 case}
\begin{figure}
\begin{center}
\includegraphics[scale=.3]{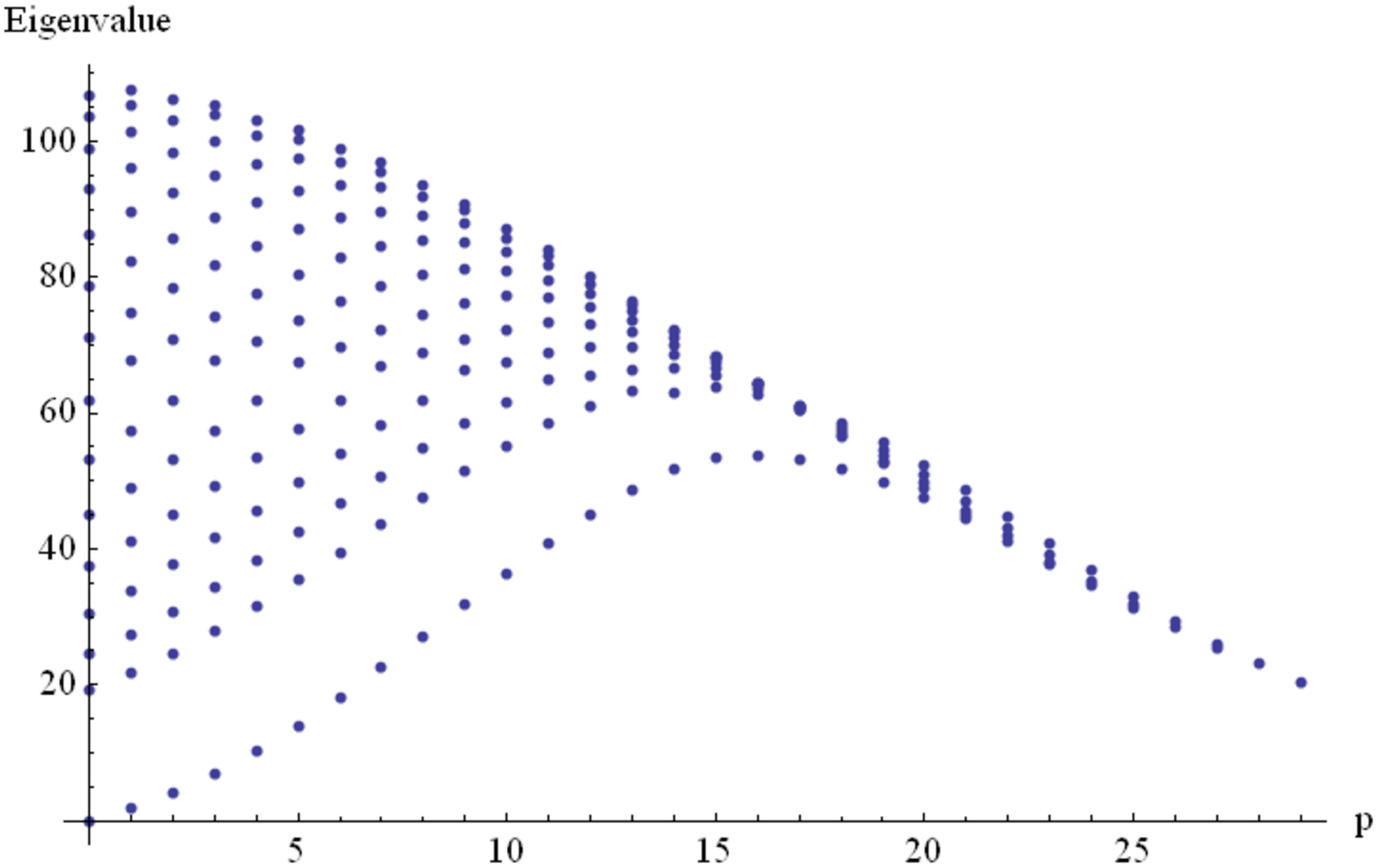}
\hfil
\includegraphics[scale=.3]{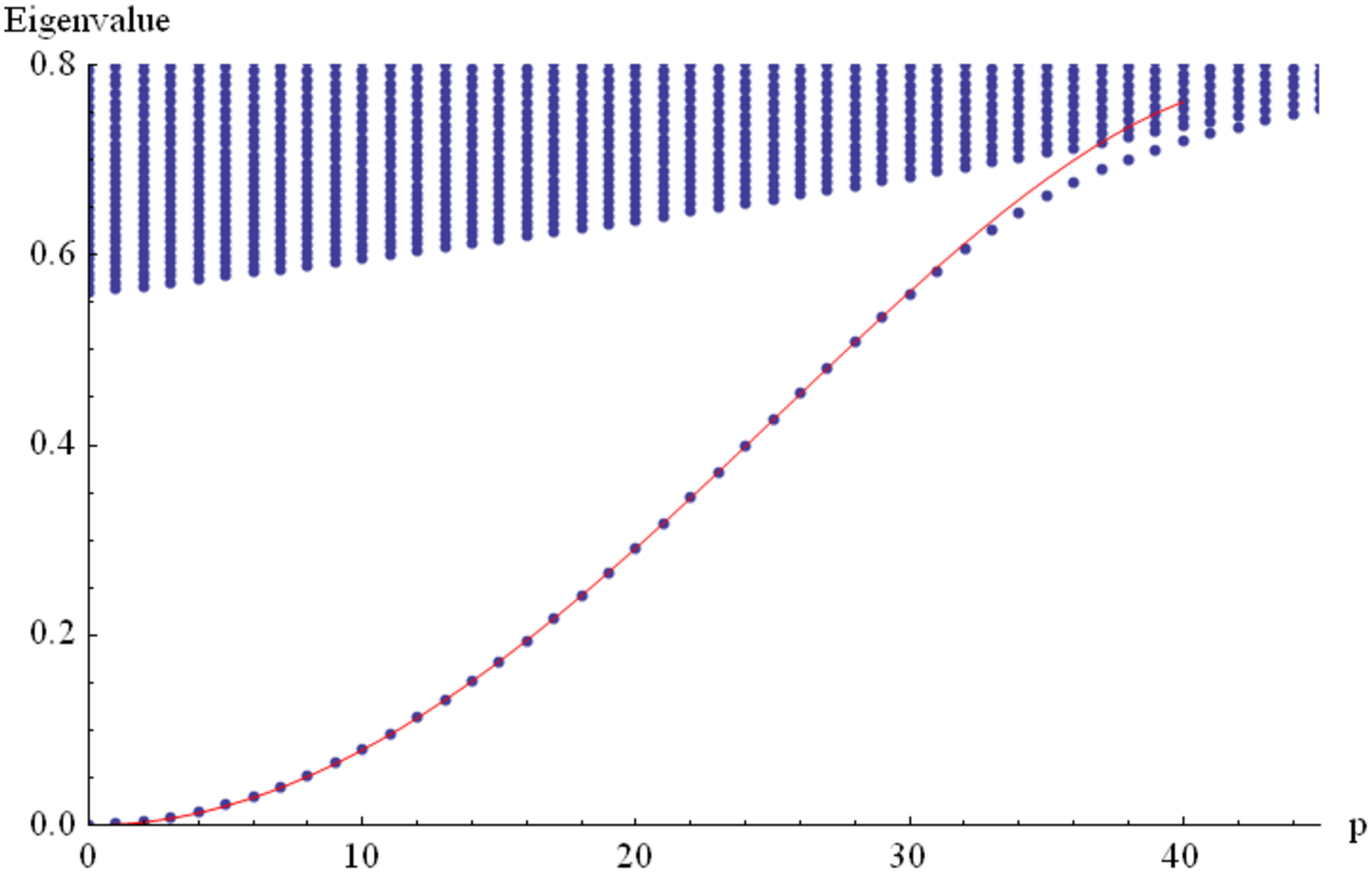}
\caption{The eigenvalue plots for $D=1$. The horizontal axis is the momentum of the ghosts.
The left figure shows the whole spectra for $L=15$, $\alpha=0.5/L^2$.
The right figure shows the low part of the spectra for $L=1500$, $\alpha=3/L^2$. 
The fitting line is $8.1\times 10^{-4} p^2 -2.1\times 10^{-7}p^4$.}
\label{fig:fig1}
\end{center}
\end{figure}

In Fig.\ref{fig:fig1}, the eigenvalues of $M^{gh}$ are plotted for two cases. 
A zero spectrum exists at $p=0$, as expected from the fact that 
there remains a translational symmetry on the 
background. 
There clearly exist a series of spectra which form a massless trajectory and 
are clearly separated from the other higher modes at low momenta.
This series can be identified with the reparametrization
ghost of the continuum theory, as explained below. 
In fact, the trajectory contains only one mode at each momentum value,
which is consistent with the result \eq{eq:spectracon} of the continuum theory.
In the left figure with $L=15$, $\alpha=0.5/L^2$, 
the trajectory looks to have a linear momentum dependence
near the origin, which contradicts \eq{eq:spectracon}. 
But, as can be seen in the right figure with $L=1500$, $\alpha=3/L^2$,
the trajectory tends to become reasonably smooth near the origin in the cases with larger $L$ and $\alpha$, 
which is consistent with \eq{eq:spectracon}. 
This is in agreement with the natural expectation 
that the continuum theory will be obtained only in large $L$ and at low momenta.

\subsection{The D=2 case}
\begin{figure}
\begin{center}
\includegraphics[scale=.3]{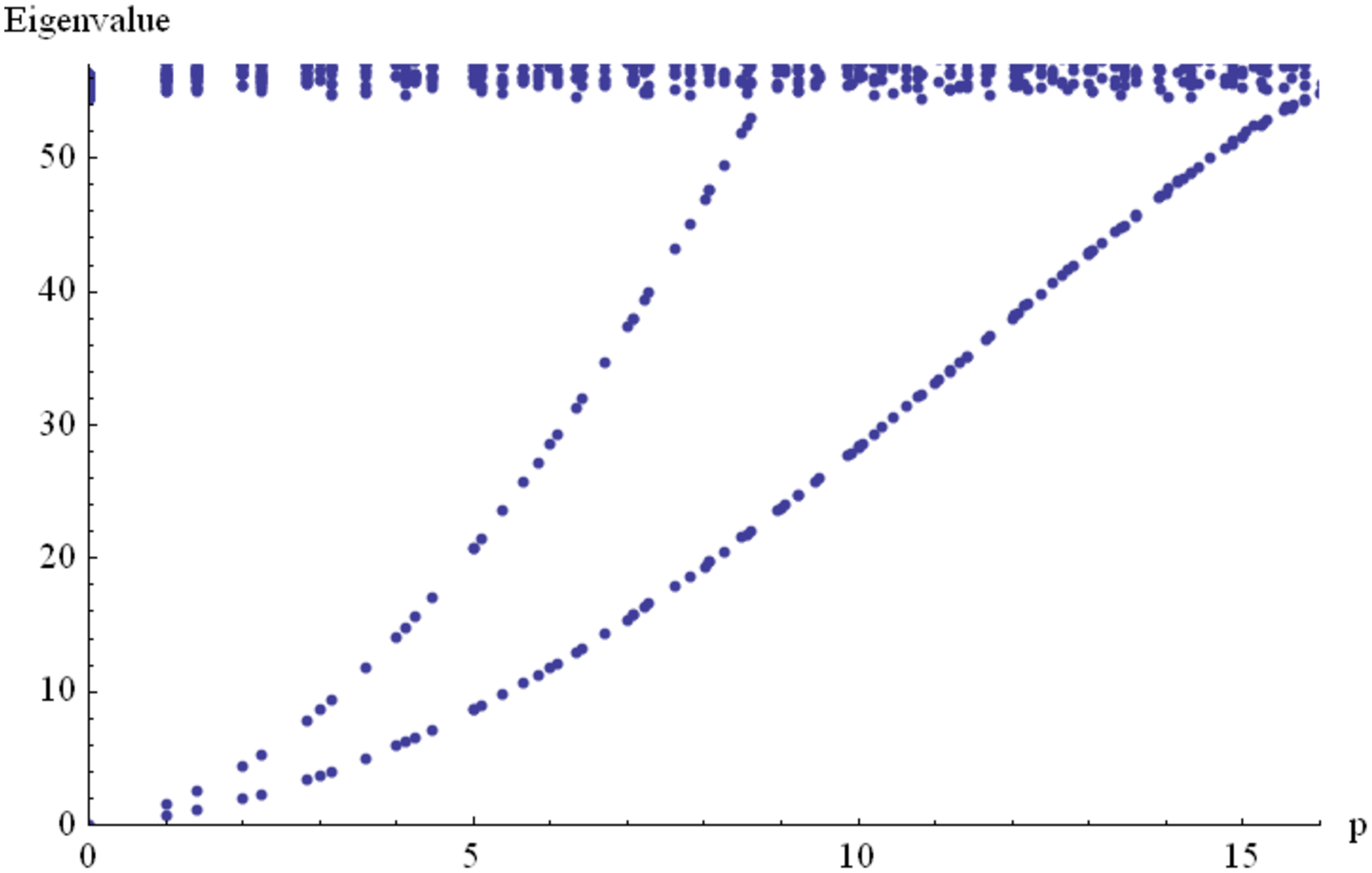}
\hfil
\includegraphics[scale=.3]{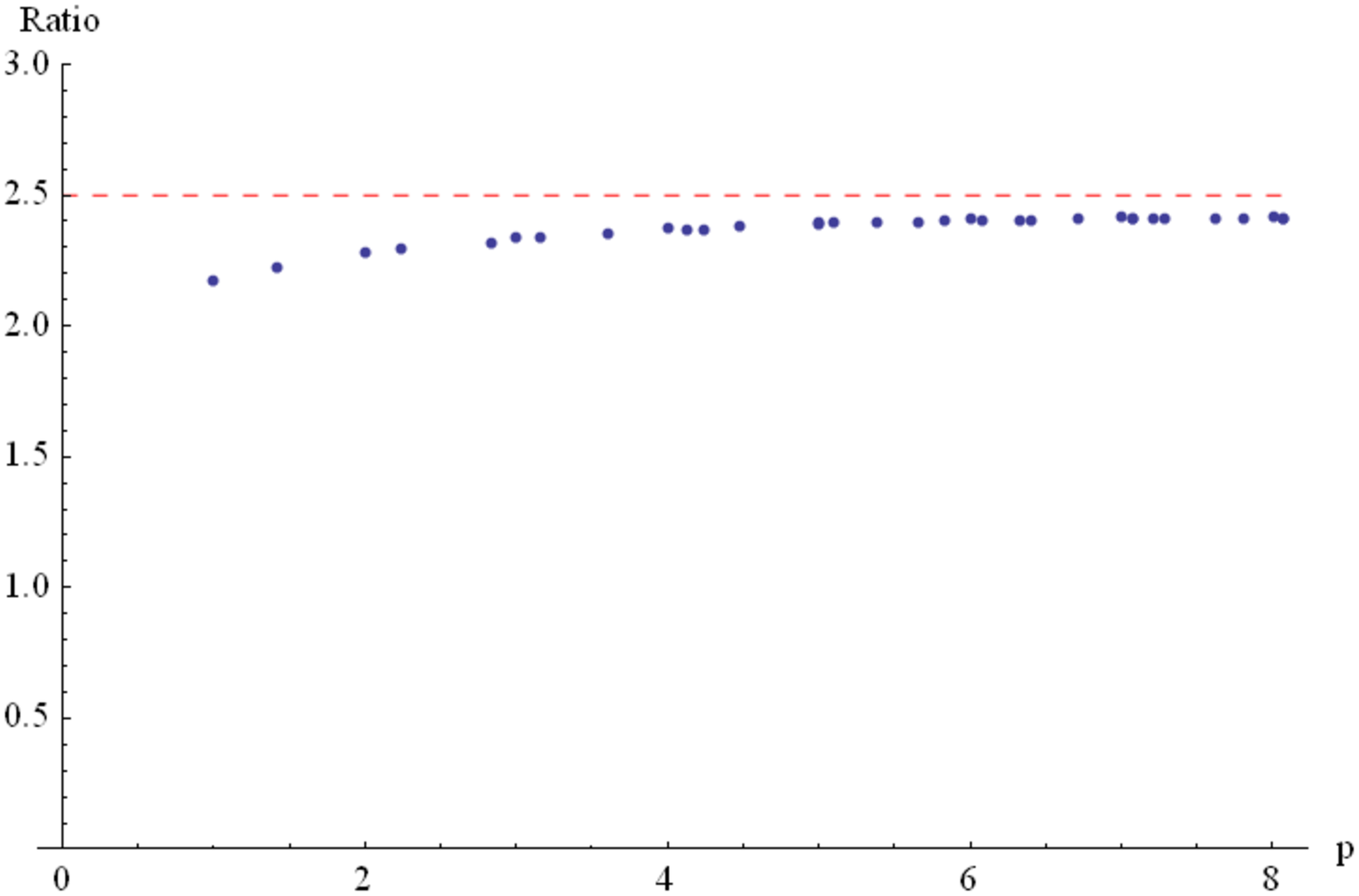}
\caption{The left figure shows the low part of the spectra for $D=2$, $L=100$, $\alpha=2/L^2$.
The right figure shows the ratios of the two trajectories.
The horizontal axis is the momentum size $\sqrt{(p^1)^2+(p^2)^2}$.}
\label{fig:fig2}
\end{center}
\end{figure}

The result \eq{eq:spectracon} of the continuum theory implies that  
there should exist two massless trajectories of spectra with no degeneracy, 
and that the ratios of the spectra in the two trajectories should be given by $\frac{20}{8}=2.5$. 
In fact, in the left figure of Fig.\ref{fig:fig2}, 
one can find that there exist two massless trajectories linked to the two zero spectra at $p=0$,
which come from the unbroken translational symmetries.
The numerical data also show that
each trajectory has only one mode at each momentum value. 
The right figure shows that the ratios of the two trajectories at each momentum
are actually in good agreement with $\frac{20}{8}$.  

\begin{figure}
\begin{center}
\includegraphics[scale=.2]{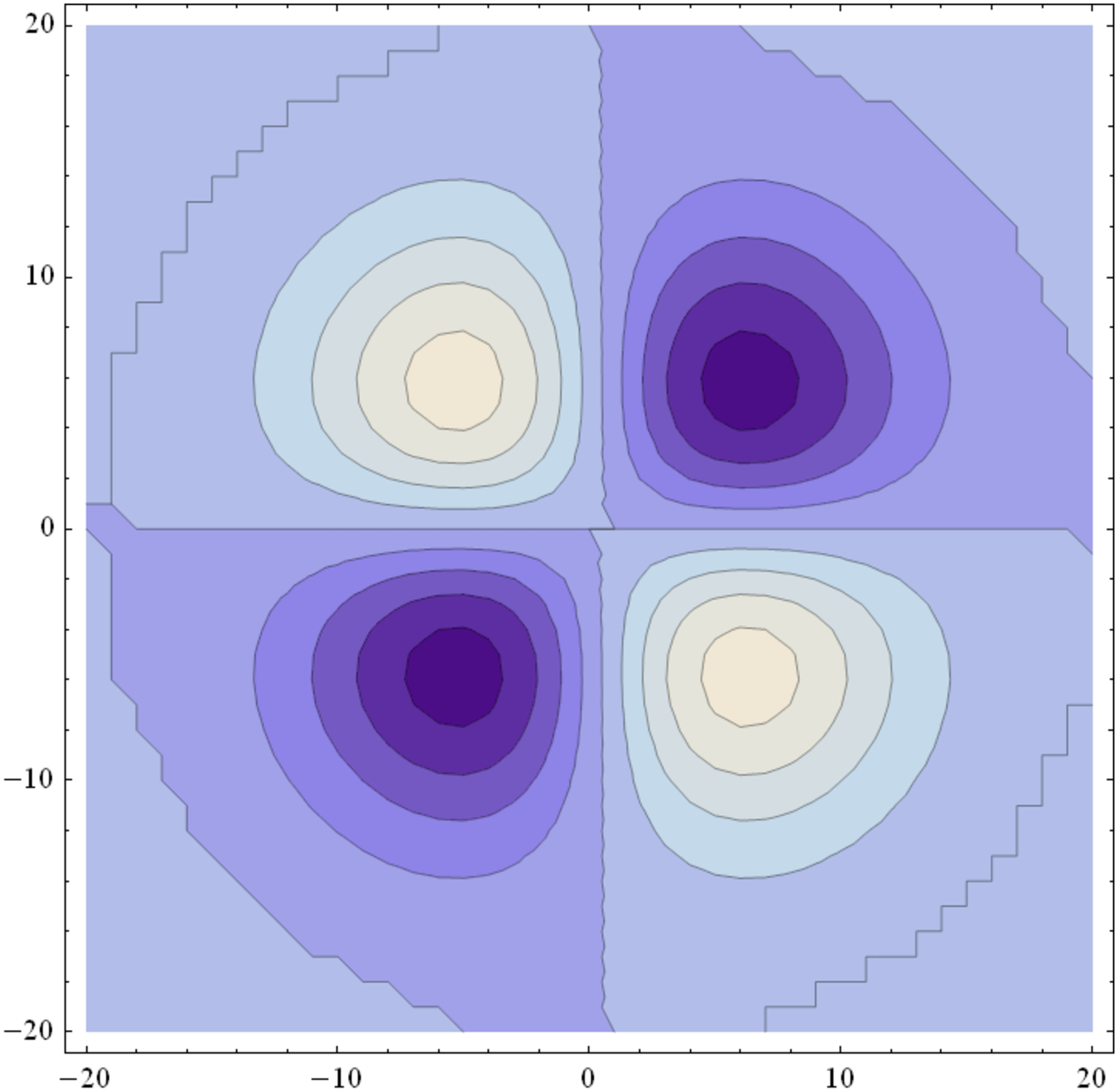}
\hfil
\includegraphics[scale=.2]{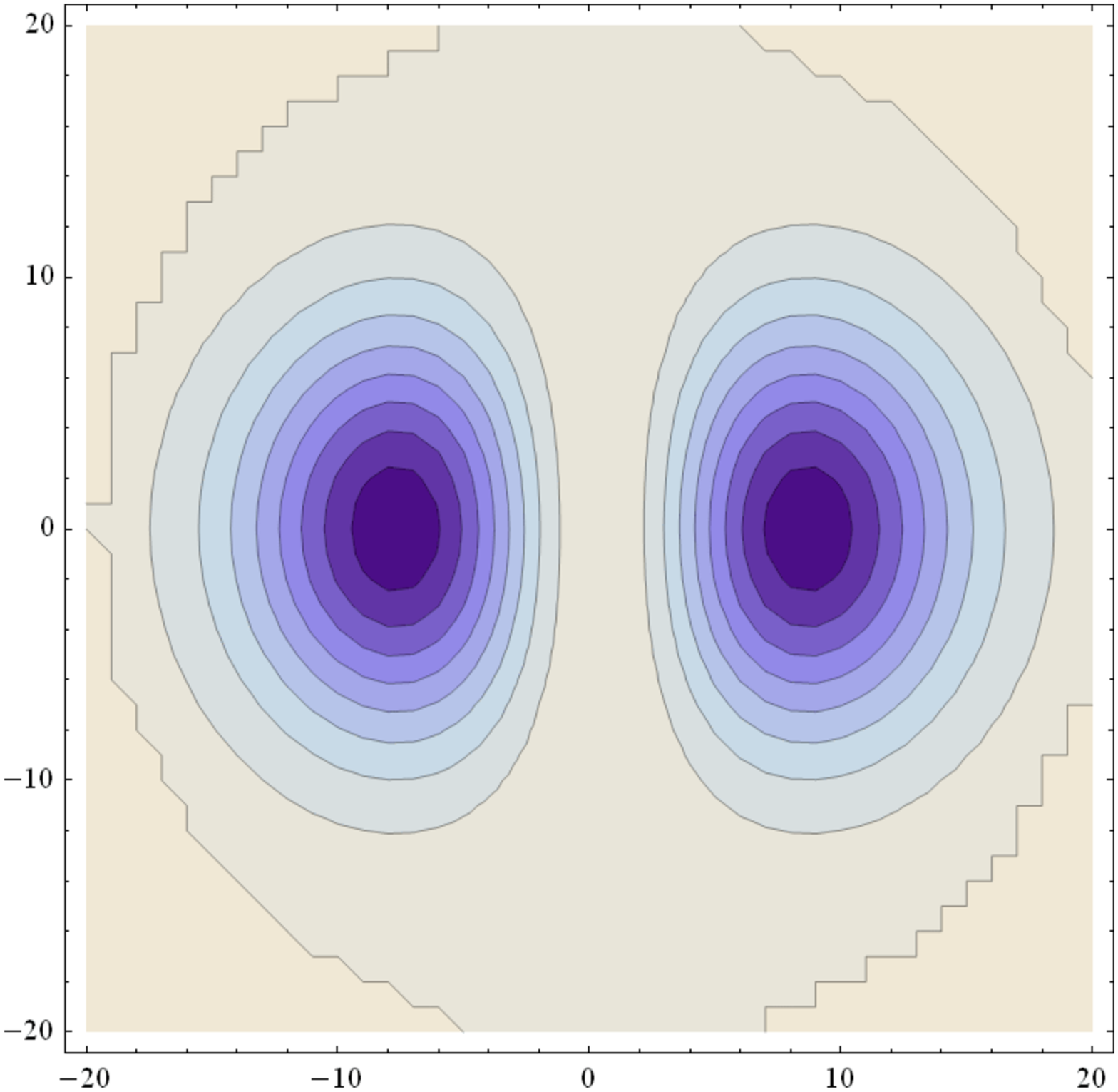}
\caption{The contour plots of $\delta K_{q,-q+p}$ for
the modes contained in the two massless trajectories 
for $D=2$, $L=20$, $\alpha=2/L^2$. The horizontal axes are $(q^1,q^2)=q$.
The left and right figures are shown for the modes 
in the lower and upper trajectories, respectively,
at the momentum $p=(1,0)$.}
\label{fig:fig3}
\end{center}
\end{figure}

It will also be a good check to see whether
the mode profiles are consistent with the continuum theory.
To do this, I follow the strategy in the previous 
works \cite{Sasakura:2007ud,Sasakura:2008pe}.
Let me define a tensor,
\be
K_{ab}=C_{acd}C_b{}^{cd}.
\ee
Under small fluctuations around $C^0$, this tensor fluctuates as
\be
\delta K_{ab}=\delta C_{acd}C^0{}_b{}^{cd}+C^0_{acd}\delta C_b{}^{cd}.
\ee
The present interest is in the fluctuations in the gauge directions $T^iC^0$.
For the gauge direction determined by an eigenvector $v$ of $M^{gh}$, $\delta K$ is
given by  
\be
\delta K^v_{ab}=(v_iT^iC^0)_{acd} C^0{}_b{}^{cd}+C^0{}_{acd} (v_iT^iC^0)_b{}^{cd}. 
\ee
In Fig.\ref{fig:fig3}, $\delta K$ is plotted for the eigenmodes contained in the two trajectories.

\begin{figure}
\begin{center}
\includegraphics[scale=.2]{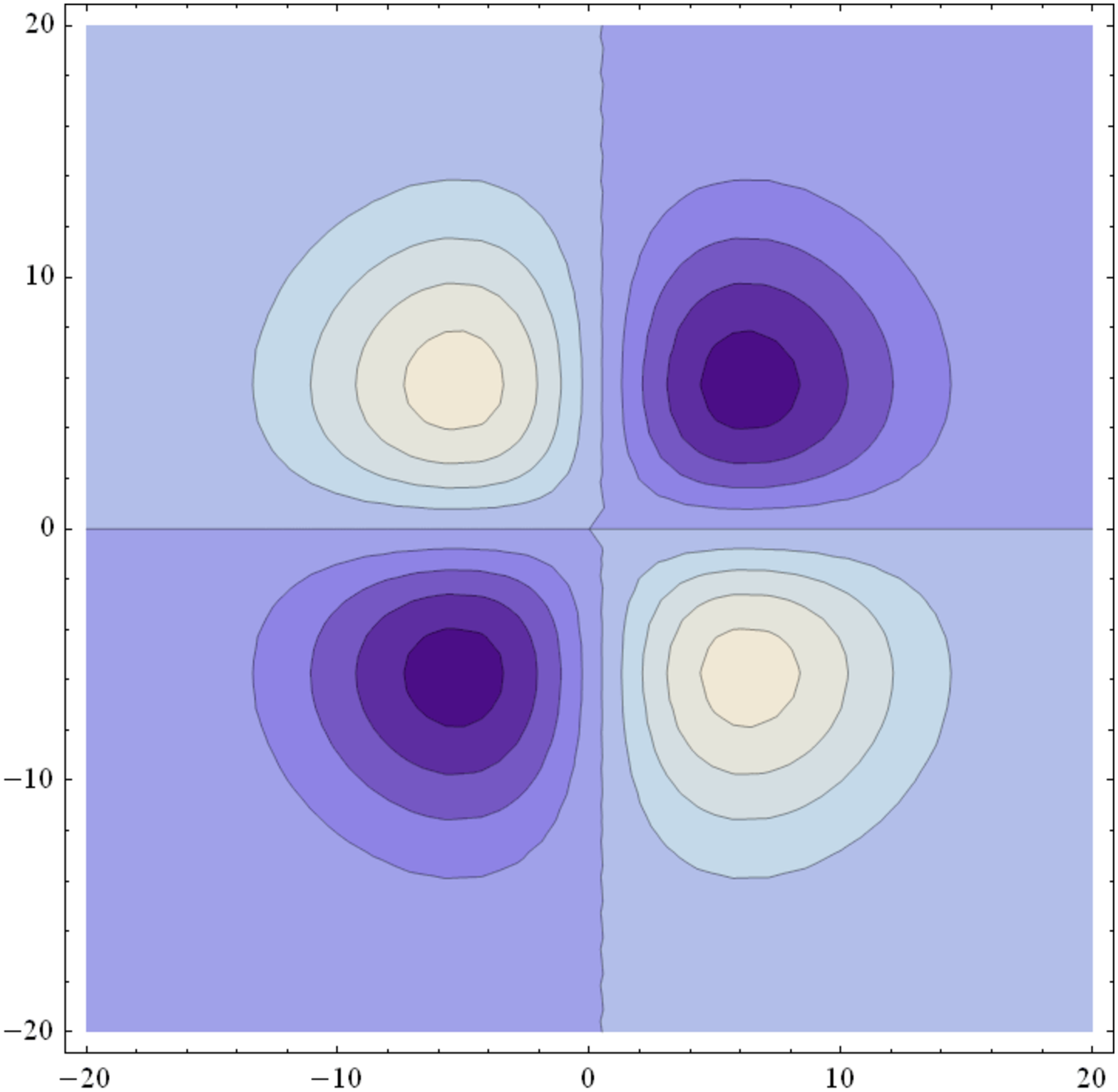}
\hfil
\includegraphics[scale=.2]{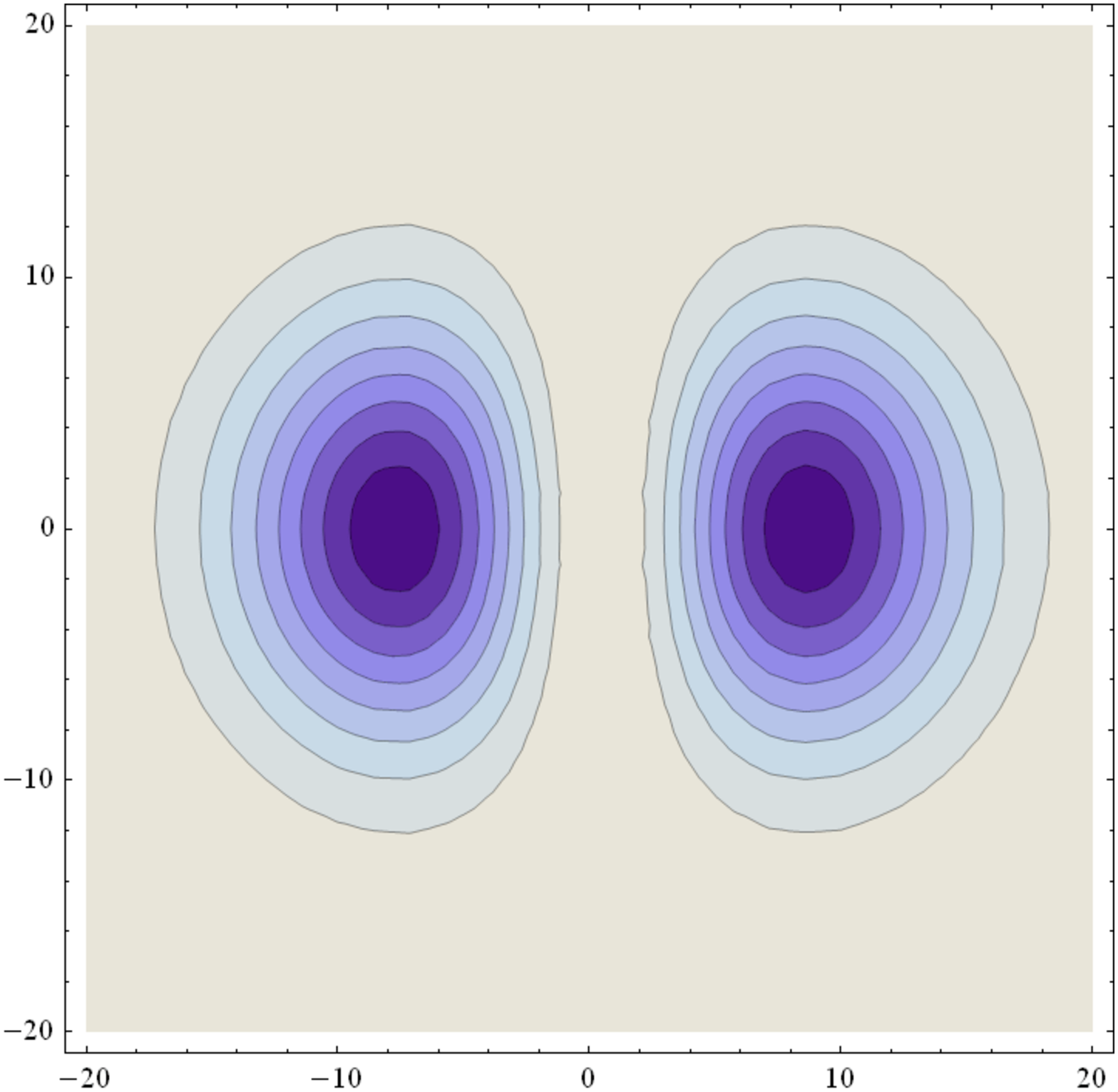}
\caption{The contour plots of $\delta K_{q,-q+p}$ for $p=(1,0)$ expected from the continuum theory. 
The left and right figures are for 
the normal ($p_\mu n^\mu=0$) and the longitudinal ($p_\mu\propto n_\mu$) modes, respectively. }  
\label{fig:fig4}
\end{center}
\end{figure}

On the other hand, the correspondence between the tensor model and 
the general relativity implies \cite{Sasakura:2007ud,Sasakura:2008pe}
\be
\delta K_{p_1p_2}=h_{\mu\nu}(p_1+p_2)\, (p_1-p_2)^\mu (p_1-p_2)^\nu \exp 
\left( -\frac{3\alpha}{4} (p_1-p_2)^2\right).
\ee
Substituting the gauge transformation $h_{\mu\nu}(p)= n_\mu p_\nu+n_\nu p_\nu$ into this expression,
one obtains the two contour plots in Fig.\ref{fig:fig4} 
for the normal ($p_\mu n^\mu=0$) and the longitudinal ($p_\mu\propto n_\mu$) modes,
respectively.
These figures are in full agreement with Fig.\ref{fig:fig3} in view  of the mode assignment
\eq{eq:spectracon}.

\subsection{The $D=3$ case}
\begin{figure}
\begin{center}
\includegraphics[scale=.3]{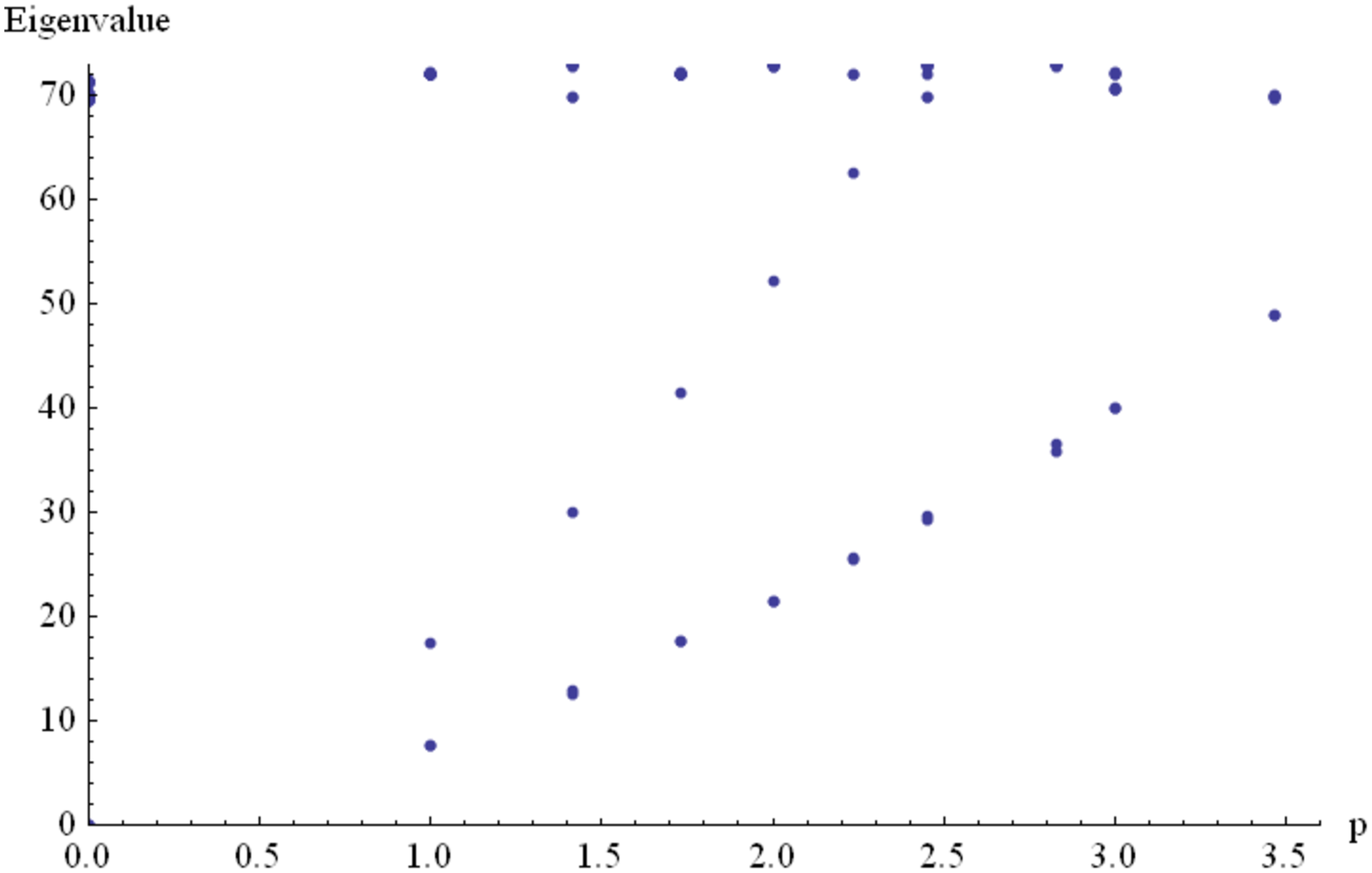}
\hfil
\includegraphics[scale=.3]{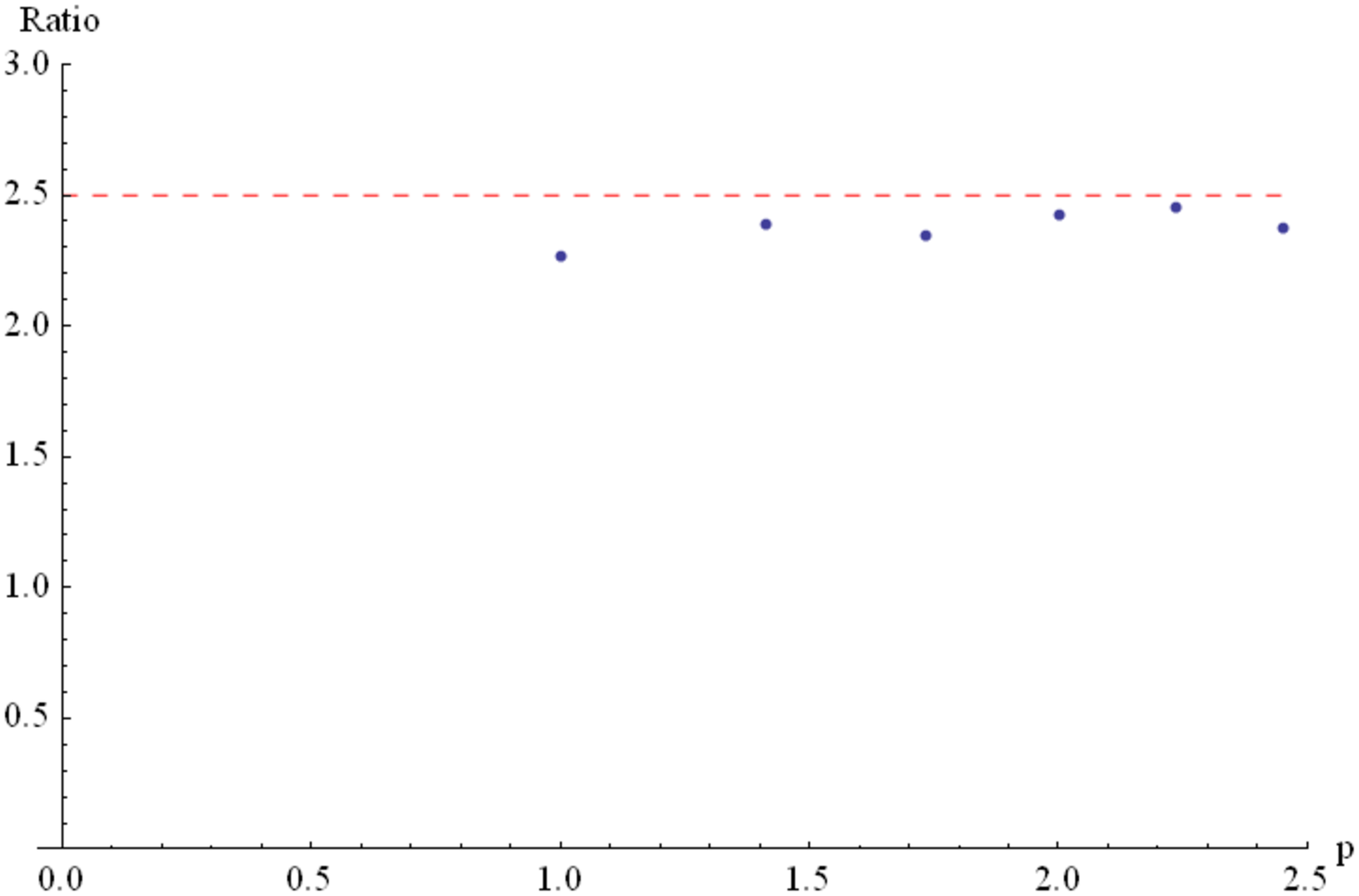}
\caption{The left figure shows the low part of the spectra for $D=3$, $L=15$, $\alpha=1.5/L^2$.
The right figure shows the ratio of the two trajectories.}
\label{fig:fig5}
\end{center}
\end{figure}

The result \eq{eq:spectracon} of the continuum theory implies that
the lower trajectory should contain two modes at each momentum.
In the left figure of Fig.\ref{fig:fig5}, the low part of the spectra for $D=3$, $L=15$, $\alpha=1.5/L^2$ 
is shown. There exists two massless trajectories, and in fact the numerical data
show that the lower trajectory contains two modes at
each momentum value. In the right figure, the ratios of the two trajectories are shown, 
which are consistent with $\frac{20}{8}$.  

\section{Summary and discussions}
In this paper, I have applied the BRS gauge fixing procedure to the tensor model, and 
have numerically analyzed the ghost kinetic term at the Gaussian backgrounds,
which correspond to the fuzzy flat spaces with arbitrary dimensions.
Then it has been found that there exist some massless trajectories of the ghost modes, 
which are clearly separated from the other higher ghost modes.
By examining the properties of the modes in these massless trajectories, it has been 
shown that these modes can be identified with the reparametrization ghosts in
the BRS gauge fixing of the general relativity. 
This means physically that the local gauge symmetry (the local translation symmetry) is emergent 
around these backgrounds in the tensor model.

Combined with the results of the previous works, this paper has shown that
the low-lying fluctuations around the Gaussian backgrounds in the tensor model  correctly generates
the general relativity, including its gauge symmetry, on the flat spaces in general dimensions.
However, this has only been shown 
in the quadratic order of the fluctuations around the backgrounds, 
but not for any of the higher non-linear terms. 
On the other hand, the general relativity (possibly with modification of the action) 
is the unique interacting field theory of the rank-two symmetric tensor field with the gauge symmetry.   
Therefore there exists a good chance for the tensor model to correctly generate also the non-linear 
terms. This should be studied in future works.
 
The above agreement between the tensor model and the general relativity including the gauge symmetry
also suggests that the quantization of the general relativity can be realized by 
that of the tensor model, and thus a kind of quantum gravity can be defined 
by the tensor model.  
There exist a lot of questions to be addressed by quantum gravity, 
the most phenomenologically interesting of which would be the cosmological 
constant problem \cite{Weinberg:1988cp}. 
In the conventional approaches, one needs an extreme fine-tuning of the cosmological constant
to stabilize a flat space from quantum corrections.
Therefore it would be interesting to study 
how the Gaussian backgrounds, which represent fuzzy flat spaces, respond to quantum corrections
in the tensor model.

So far, the correspondence between the tensor model and the general relativity has been shown only 
for a limited class of tensor models, which have the Gaussian solutions.
It is also an interesting future problem to investigate
whether such correspondence holds in more general tensor models.

\vspace{.5cm}
\section*{Acknowledgments}
The author would like to thank T.~Kugo for the knowledge about the references on
the BRS gauge fixing in the general relativity and those on the viewpoints of gauge symmetry
as spontaneously broken symmetry.
The author was supported in part by the Grant-in-Aid for Scientific Research 
No.18340061(B) from the Japan Society for the Promotion of Science (JSPS).


\end{document}